\documentclass[sn-mathphys-ay]{sn-jnl}

\usepackage{graphicx}%
\usepackage{multirow}%
\usepackage{amsmath,amssymb,amsfonts}%
\usepackage{amsthm}%
\usepackage{mathrsfs}%
\usepackage[title]{appendix}%
\usepackage{xcolor}%
\usepackage{textcomp}%
\usepackage{manyfoot}%
\usepackage{booktabs}%
\usepackage{algorithm}%
\usepackage{algorithmicx}%
\usepackage{algpseudocode}%
\usepackage{listings}%
\usepackage{natbib}%
\setcitestyle{round}%

\usepackage{enumitem}

\usepackage{lmodern}
\usepackage{anyfontsize}

\graphicspath{{figs/}}

\begin{document}

\title[Article Title]{GLARE: Guided LexRank for Advanced Retrieval in Legal Analysis}

\author[1]{\fnm{Fabio} \sur{Gregório}}\email{fabio.gregorio@aluno.cefet-rj.br}

\author[1,3]{\fnm{Rafaela} \sur{Castro}}\email{rafaela.nascimento@serpro.gov.br}
\equalcont{These authors contributed equally to this work.}

\author[1]{\fnm{Kele} \sur{Belloze}}\email{kele.belloze@cefet-rj.br}
\equalcont{These authors contributed equally to this work.}

\author[2]{\fnm{Rui} \sur{Pedro Lopes}}\email{rlopes@ipb.pt}
\equalcont{These authors contributed equally to this work.}

\author*[1]{\fnm{Eduardo} \sur{Bezerra}}\email{ebezerra@cefet-rj.br}
\equalcont{These authors contributed equally to this work.}

\affil[1]{\orgdiv{Computer Science Department}, \orgname{Federal Center for Technological Education (CEFET/RJ)}, \orgaddress{\city{Rio de Janeiro}, \state{RJ}, \country{Brazil}}}

\affil[2]{\orgdiv{CeDRI, SusTEC}, \orgname{Instituto Politécnico de Bragança (IPB)},\\ \orgaddress{\city{Bragança}, \country{Portugal}}}

\affil[3]{\orgdiv{Technology Department}, \orgname{Serviço Federal de Processamento \\de Dados (SERPRO)}, \orgaddress{\city{Rio de Janeiro}, \state{RJ}, \country{Brazil}}}


\abstract{The Brazilian Constitution, known as the Citizen's Charter, provides mechanisms for citizens to petition the Judiciary, including the so-called special appeal. This specific type of appeal aims to standardize the legal interpretation of Brazilian legislation in cases where the decision contradicts federal laws. The handling of special appeals is a daily task in the Judiciary, regularly presenting significant demands in its courts. We propose a new method called GLARE, based on unsupervised machine learning, to help the legal analyst classify a special appeal on a topic from a list made available by the National Court of Brazil (STJ). As part of this method, we propose a modification of the graph-based LexRank algorithm, which we call Guided LexRank. This algorithm generates the summary of a special appeal. The degree of similarity between the generated summary and different topics is evaluated using the BM25 algorithm. As a result, the method presents a ranking of themes most appropriate to the analyzed special appeal. The proposed method does not require prior labeling of the text to be evaluated and eliminates the need for large volumes of data to train a model. We evaluate the effectiveness of the method by applying it to a special appeal corpus previously classified by human experts.}


\maketitle

\section{Introduction}\label{sec:introduction}

The Brazilian Federal Constitution of 1988 promoted an expansion of means for citizens to seek their rights, resulting in a significant increase in cases in the Judiciary. If, on the one hand, the constitutional text brought advances in guaranteeing citizens' rights, it also brought the challenge of quickly meeting society's demands. According to official data from the National Council of Justice (CNJ) in Brazil, the Judiciary ended 2022 with 81.4 million processes in progress awaiting a definitive solution~\citep{cnj:online}.

The Constitution provides means by which a citizen may provoke action by the Judiciary if they feel someone has violated their rights. One of these means is the so-called appeal, which is the proper means to challenge a judicial decision aiming at its re-examination, to try to obtain, in the same procedural relationship, the reform, invalidation, clarification, or integration of the judgment. In a general sense, appeal is the power to review a decision by the same judicial authority or by another hierarchically superior party, aiming to obtain its reform or modification~\citep{Constitu51:online}.

A specific type of appeal provided for in the Constitution is the special appeal. This appeal fulfills the purpose of standardizing the legal interpretation of the Brazilian rules. The National High Court of Brazil (STJ) judges this appeal in situations of cases that were decided, solely or lastly instance, by the Federal Regional Courts or by the courts of the States, of the Federal District and Territories. The judgments are made when the appealed decision: contravenes a treaty or federal law, or denies them validity; judge a contested act of local government valid in the face of federal law; give federal law an interpretation divergent from that assigned to it by another court~\citep{Constitu51:online}.

The STJ set up legal mechanisms to accelerate the judicial process and improve the court workflow. One of these mechanisms is the so-called \textit{repetitive appeal}, set up by STJ with Law 11,672/2008. With application in the judgment of special appeals that deal with the same legal controversy, the repetitive appeal system establishes that, for similar demands, the STJ will judge sampling processes selected ~\citep{CNJServi8:online}.

For each controversy sent to the STJ through a special appeal, a legal thesis will be defined and shown that must be applied to all appeals in which an identical law issue is discussed. The lower courts are responsible for following the Civil Procedure Code (CPC) when evaluating the requirements necessary for the admissibility of each special appeal. These courts are also responsible for investigating whether there is a legal thesis published by the STJ that should be applied to the admitted appeal.

The above-described procedure to serve incoming special appeals is a typical every-day case within the Brazilian Judicial Branch. Classifying a special appeal on a theme is a complex procedure whose main task is to comprehensively analyze the content of the incoming appeal and compare the facts presented with the themes present in STJ's repetitive appeals system. The aim is to find eventual similarities with an existing theme. This comparison is necessary to manage each incoming appeal correctly, which can be a referral to the STJ as a case to be judged in the repeated appeals system, suspend the process until a thesis on a given theme is defined, or return the case to its original court so that the understanding already established in a thesis previously published by the STJ is respected.

There are three essential aspects to highlight about repetitive themes. First, each repetitive theme corresponds to a small text (about 36 words per theme), typically consisting of a single paragraph without a rigidly defined structure. This aspect is relevant because when we classify a special appeal in a repetitive theme, we deal with texts of a very different order of magnitude (a special appeal has about 3,980 words).

Secondly, the collection of repetitive themes is periodically increased due to changes in current legislation and new societal demands. This aspect directly influenced our adoption of an unsupervised learning approach. The scarcity of training data directly affects the performance of a supervised learning approach.

Lastly, there may be similar themes, as shown in the following examples, in which both themes refer to the same law, the same article, and the limitation period:
\begin{itemize}
        \item Theme 568: ``The system for counting intercurrent prescriptions is discussed (presence creation after filing the action) provided for in art. 40 and paragraphs of the Enforcement Law Tax (Law No. 6.830/80): What are the obstacles to the course of the statute of limitations? Prescription provided for in art. 40, from LEF.''
        \item Theme 569: ``The system for counting intercurrent prescriptions is discussed (presence creation after filing the action) provided for in art. 40 and paragraphs of the Enforcement Law Tax (Law No. 6.830/80): if the absence of a notice from the Public Treasury regarding the order that determines the suspension of tax execution (art. 40, § 1) contradicts the decree of the intercurrent prescription.''
    \end{itemize} 
    
This aspect influenced our choice to treat the problem as a legal case retrieval. Legal case retrieval refers to the process of identifying and retrieving relevant legal cases from a database based on a query or specific criteria. This task involves searching through vast collections of legal documents to find cases that are pertinent to a particular legal issue, question, or context~\citep{SANSONE2022101967}. Due to the subtle details that can differentiate one theme from another, we understand that it is convenient to provide a small list of themes as an output of our method so that the human expert can choose the appropriate theme instead of offering a single theme.

The characteristics of special appeals and repetitive themes require the human legal analyst to have good synthesis skills. They will have to deal with documents exceeding four thousand words each and match them with a text of about forty words. A common scenario that the analyst needs to deal with is the low availability or even the total absence of a special appeal previously classified into a theme, similar to the incoming appeal under analysis.

According to official data from 2021, this year alone, more than 50,000 special appeals were filed with the STJ~\citep{cnj:online}. The National Council of Justice (CNJ) maintains a statistical dashboard for the Judiciary, but there is no data on the average time required to perform specific tasks. At the Regional Federal Court of the 2nd Region (TRF2), the vice president decides on the admissibility of special appeals, which legal analysts handle. By analyzing data from TRF2, we found out that the average time between the admission of a special appeal by the vice presidency and a decision on the appeal is 113.47 days. A tool that could make the process of assigning a special appeal faster would be important to reduce this processing time.

This article presents an effective unsupervised machine learning method to support a legal analyst in finding a match between an incoming special appeal and an element in the set of repetitive themes. Our approach consists of first summarizing the special appeal given as input to highlight its most relevant informational points. In the second step, we evaluate the similarity between that summary and each available theme. Finally, we generate an ordered list of the themes most similar to the summary and the special appeal given as input. As part of our method, we propose an algorithm for text summarization.

Initially, we used the TRF2’s solution as a baseline. This solution uses an open-source distributed search engine called Elasticsearch\footnote{https://www.elastic.co/guide/en/elasticsearch/reference/current/index-modules-similarity.html}. This baseline evaluates the similarity between the new special appeal and existing appeals in the historical database, which were previously classified into a theme. A theme is suggested for the new special appeal being analyzed according to the degree of similarity calculated.

To obtain a more robust validation, additionally, we evaluated two supervised learning models. The first one uses XGBoost, which is widely considered a state-of-the-art technique for supervised classification~\citep{Chen_2016}. As a second supervised baseline, we used Logistic Regression due to its well-established calibration properties and the high confidence it provides in its predictions\footnote{https://scikit-learn.org/stable/modules/calibration.html}. With the corpus used in the experiments, our method was able to suggest the correct topic for approximately 76\% of the special appeals. This result was much higher than the approach adopted by TRF2, which achieved approximately 35\% for the same dataset. Compared to supervised learning models, the result was surprising in critical cases. In scenarios with little labeled data on specific topics, our method achieved a rate of 72\%, while the logistic regression model achieved approximately 60\%, and XGBoost achieved approximately 29\%.
In a scenario in which legal themes were not represented in the training data, and only in the test data, the supervised models were unable to suggest the correct topic in any case. While our model achieved a rate of approximately 70\%.

The main contributions of this paper are twofold: (i) We assemble a pair of corpora, one about special appeals and the other about repetitive themes, and make them publicly available for the research community. (ii) We propose an unsupervised method to match elements between the sets of repetitive themes and special appeals.

The rest of this article is organized as follows. In Section~\ref{sec:background}, we present background content to give context to subsequent sections. We present related work in Section~\ref{sec:related:work}. Section~\ref{sec:corpus} presents the legal datasets adopted in this work. Section~\ref{sec:methodology} presents an overview of the proposed method. In Section~\ref{sec:experiments}, we describe the experiments that aim to verify the effectiveness of the proposed method. Section~\ref{sec:conclusion} presents final remarks and future work.

\section{Background}
\label{sec:background}
This section establishes the theoretical foundation for the unsupervised model used in this study to classify special appeals. Section~\ref{sec:background:summarization} provides an overview of text summarization techniques, categorizing them into three main approaches. It highlights the risks of hallucination in abstractive methods and details extractive techniques like topic-based and centrality-based summarization, with emphasis on the BERTopic model and the LexRank algorithm. Section~\ref{sec:background:lexrank} provides an overview of Lexrank, a graph-based algorithm that served as a model for the summarization algorithm we developed in this project.

\subsection{Summarization Techniques}
\label{sec:background:summarization}

Summarization techniques can be categorized into information fusion, abstractive, and extractive~\citep{mehta2019extractive}. Information fusion techniques present characteristics of extractive and abstractive techniques. In abstractive techniques, the focus is on rewriting the text in the most precise form. Models that generate abstractive summarization are more susceptible to a phenomenon known as ``hallucination'', in which terms and structures incompatible with the original text appear in the generated summary and distort the meaning of the source text~\citep{huang2023factual}. In extractive techniques, sentences are extracted from a document or set of documents in the same form they originally appeared. There are three main variations in extractive techniques:
\begin{itemize}
     \item Topic-based—Focus on assigning relative importance to words or sentences. The system can assign weights based on the frequency of terms. By storing information about frequencies, selecting the most important words or sentences is possible.
     \item Centrality-based—They try to find the sentence that best is the general content of the document. A ranking is set up; the sentence considered most important is the one whose content is shared with the greatest number of other sentences in the document.
     \item General summary—Treats summarization as an optimization problem, trying to find which subset of sentences best is the document. It does not focus on individual sentences.
\end{itemize}

In this project, we consider two different approaches to summarizing text and computing their performance:
\begin{itemize}
    \item Topic-based—We evaluate the BERTopic technique~\citep{grootendorst2022bertopic}. This technique uses pre-trained neural network models and clustering algorithms to generate groups of documents with semantic similarity from a corpus. As a group is generated, all documents in the cluster are combined into a single document. A modified version of the TF-IDF algorithm, called class-based TF-IDF (c-TFIDF), is applied to this document. This generates, at the cluster level, something called a bag of words representation, in which the frequency of each word in the cluster can be found. The most important words are extracted and will compose the representative topic of the cluster, and consequently, the topic that represents each document that participated in the formation of the cluster. In the BERTopic library, you can perform Guided Topic Modeling or Seeded Topic Modeling, which consists of a collection of techniques that guide the topic modeling approach by defining several initial topics to which the model will converge. The objective is to influence the words that will make up the cluster topic. In our project, we evaluated the use of the set of words belonging to repetitive themes to guide the BERTopic model in the process of discovering topics representative of special appeals.
    \item Centrality-based—We evaluate the LexRank algorithm and a variation of it which we call Guided LexRank, which is a key component of our method GLARE.
\end{itemize}

\subsection{LexRank}
\label{sec:background:lexrank}
\begin{figure}[!ht]
	\centering
	\includegraphics[width=0.5\textwidth]{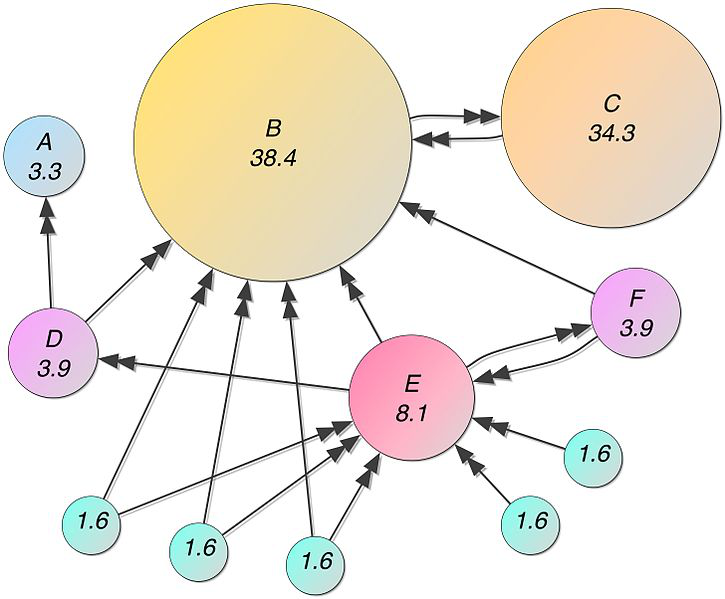}
	\caption{PageRank is the basis for other graph-based algorithms such as LexRank. Source: Mehta, Parth and Prasenjit Majumder. From Extractive to Abstractive Summarization: A Journey (2019, p.12)~\citep{mehta2019extractive}}
	\label{fig:pagerank}
\end{figure}

Graph theory has applicability in several areas of knowledge since the objects of a graph are abstractions that can represent countless types of entities. \textbf{Pagerank} is a graph-based algorithm that uses the hyperlinks structure of web pages to create a ranking based on reputation. Taking a graph $G$ as an abstraction, web pages are represented as vertices $v \in V$ and the hyperlinks between pages are edges $e \in E$. A vertex receives a weight (reputation) that is associated with relationships with other vertices (Figure ~\ref{fig:pagerank}). The basic idea is that pages with high reputations are linked to other pages that also have high reputation~\citep{DBLP:books.sp.Aggarwal22}.

The central idea of \textbf{LexRank} is to represent text as a graph, where nodes represent sentences and edges represent the similarity between sentences. The algorithm aims to summarize by extracting the most relevant sentences from a text. A minimum similarity threshold is established, the algorithm calculates the similarity between all pairs of sentences in the text, and the computed values determine the \textit{degree of centrality} of each sentence. Sentences similar to a greater number of other sentences have a higher degree of centrality. At the end of the process, the algorithm groups the sentences with a higher degree of centrality to compose the summary of the original text according to the desired summary size ~\citep{Erkan_2004}.

\section{Related work}
\label{sec:related:work}

Assigning subjects, grouping similar documents, and classifying documents are every-day tasks conducted in artificial intelligence projects related to the legal area. Typically, these tasks involve summarization generated in diverse ways.

~\citet{vianna2023topic} propose a solution based on topic discovery techniques to analyze, summarize, and classify legal documents. They use datasets from the Brazilian legal system analyzed by jurists to confirm their approach. The proposed method is divided into three phases:
 \begin{enumerate}[label=(\roman*)]
     \item Filtering—According to criteria defined by legal experts, regular expressions are used to select a subset of an ample collection of documents.
     \item Preprocessing—Removes punctuation, white spaces, and other defined terms.
     \item Discovery—is the central phase of the method in which four different techniques were implemented and evaluated.
\end{enumerate}
This work is related to ours in that we seek the theme intrinsic to the special appeal. Once this theme has been discovered, we compare it with the different themes and finally classify the resource into the most proper theme. However, in this related work, after synthesizing the texts, human experts analyze the themes generated to attest to their quality. In our approach, we calculate the similarity between texts. Because we use documents previously labeled by experts, we can check the quality of the summaries using known metrics. In this way, we minimize any bias that could interfere with the qualification of the abstracts generated. Another point of differentiation is that among the sets of documents used by the authors, there is a considerable number of documents with an average size of 220 words. In comparison, the average corpus size we used was approximately four thousand words. Still, as a difference, in our experiments, we sought to compare the BERTopic modeling adopted by ~\citet{vianna2023topic}, with graph-based summarization, the latter being the focus of our method.

\citet{10.1007/978-981-99-0085-5_46} use machine learning to summarize documents and paraphrase them into more straightforward language in work related to Indian legal precedents. The proposed method is a combination of two summarization techniques. LexRank~\citep{Erkan_2004}, an unsupervised graph-based method, is initially applied to the document to reduce it to a length corresponding to 35\% of the original size. The second step uses the PEGASUS neural language model~\citep{zhang2020pegasus} for text summarization. 

PEGASUS is based on the transformer~\citep{vaswani2023attention} architecture. It follows the encoder-decoder pattern, where the encoder processes the original text to extract contextual representations while the decoder generates the summary based on these representations. The encoder comprises several layers of encoding, while the decoder also has several layers to generate a token-by-token summary. A linear layer is applied to map the decoder outputs to the vocabulary, allowing the prediction of the following words in the summary. In the method proposed by~\citet{10.1007/978-981-99-0085-5_46}, the extractive summary generated by the LexRank model is divided into sentences and fed to the PEGASUS model. The model then derives paraphrases for each sentence, making it simpler to understand. The sentences are then merged to provide a final abstract summary.

Our work is related to~\citet{10.1007/978-981-99-0085-5_46}, who used the graph-based LexRank algorithm to create the legal document summary. However, as a distinguishing feature of our method, we modified the original LexRank algorithm, seeking a way to guide it in the summarization process and thus improve the generated summary.

~\citet{10.1007/978-981-19-0019-8_47} compare summary techniques and apply them to legal documents. The work method covers the stages of Dataset Acquisition, Data Cleaning, Text summarization, Summary evaluation, and Comparative analysis. In the extractive summarization stage, the following techniques are compared: LexRank, TextRank, Reduction, Luhn, Edmundson, Latent Semantic Analysis, and SumBasic. The results show that graph-based approaches such as LexRank and TextRank perform better than approaches based on frequency. Similarly to~\citet{10.1007/978-981-19-0019-8_47}, we attest to the good performance of the graph-based approach but make a comparison with methods that implement topic modeling.

\citet{10.1007/s10506-022-09319-6}  propose an approach to solving a case law task presented in the Legal Information Extraction and Entailment (COLIEE) competition~\footnote{\url{https://sites.ualberta.ca/~rabelo/COLIEE2020/}.}. The task is to find relevant legal cases given an input query case. The data for the task is based on the Federal Court of Canada case law. 

In the first step, the goal is to extract from the corpus the candidate cases that support the decision of the query case. In the second step of the task, given a trio including a query case, a decision of the query case, and a list of candidate paragraphs from the supporting legal documents in the previous phase, this phase aims to identify one or more paragraphs from the supporting documents cases that imply the decision of the query case. 

Both the query and the documents in the corpus are texts that average about 3,000 words. Similarly to our work,~\citet{10.1007/s10506-022-09319-6} reduce the long text into paragraph-like components using a graph-based algorithm to solve the task. In~\citet{10.1007/s10506-022-09319-6} approach, the text summary weakly labels the evaluated data. The labeled data sends the information as input to a pre-trained BERT model. In addition to the result produced by the BERT model, and similar to what we do in our approach,~\citet{10.1007/s10506-022-09319-6} submit the texts to the BM25 algorithm and use the algorithm's output value as a measure of similarity.

\citet{10.1007/s10506-022-09319-6} apply a combined score between the result produced by the BERT model and BM25 to retrieve relevant legal cases for a given query case. The proposed approach presents the relevant documents and the paragraphs supporting the choice.

\section{Corpora of special appeals and themes}
\label{sec:corpus}

To develop this work, we assembled two related corpora. The first is a corpus of documents corresponding to special appeals extracted from non-confidential legal processes within Brazil's federal justice scope. This corpus has documents corresponding to 7,967 special appeals. Table~\ref{tab:statisticsCorpus} presents a statistical summary of this corpus.

\begin{table}[htb]
\caption{Statistics of the corpus used in this work.}
\begin{tabular}{@{}lr@{}}
\toprule
\textbf{Description} & \textbf{Value} \\
\midrule
Number of documents    & 7,967   \\
Average words per document    & 4,672.48   \\
Median words per document   & 3,980 \\ 
Minimum words per document & 92 \\
Maximum words per document & 67,944 \\
Total size (in Mb) & 262 \\
\botrule
\end{tabular}
\label{tab:statisticsCorpus}
\end{table}

Section~\ref{sec:introduction} exemplifies what we refer to in this work as \textit{themes}, which are short documents referring to legal controversies. These documents form a second corpus, and Table~\ref{tab:statisticsThemes} presents statistical summaries of the themes corpus.
\begin{table}[htb]
\caption{Statistical data on the themes considered in this project.}
\begin{tabular}{@{}lr@{}}
\toprule
\textbf{Description} & \textbf{Value} \\
\midrule
         Number of themes & 190 \\
         Average number of words per theme & 43.3 \\
         Median words per theme & 36 \\
         Minimum words per theme & 6 \\
         Maximum words per theme & 300 \\
         Total size (in Kb) & 56 \\
\botrule
\end{tabular}
\label{tab:statisticsThemes}
\end{table}

\begin{figure}[htb]
    \centering
    \includegraphics[width=0.95\textwidth]{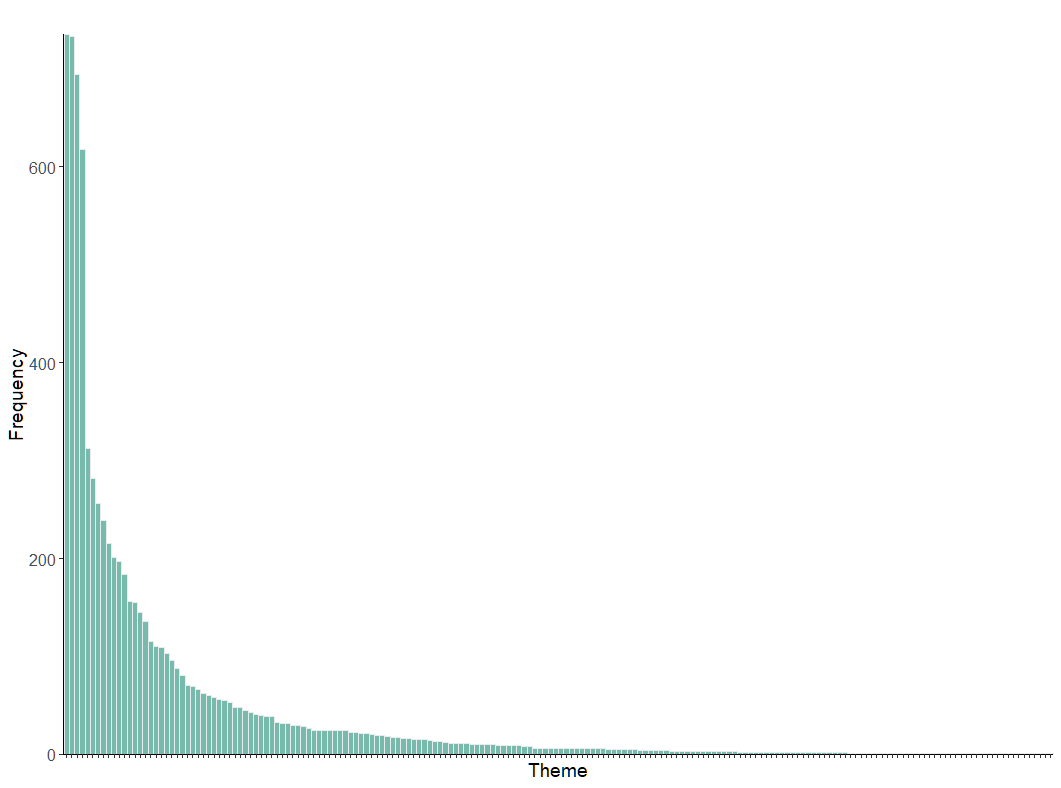}
    \caption[Histogram]{Distribution of themes in the dataset}
    \label{fig:histogram:themes}
\end{figure}

The documents that make up the corpus of special appeals were collected through public access throughout 2023. All documents were labeled with a theme given by a human expert. The documents in the special appeals corpus are distributed non-uniformly across the 190 documents that make up the thematic corpus. The histogram in Figure~\ref{fig:histogram:themes} shows this distribution. 

\section{Methodology}
\label{sec:methodology}

Our goal in this work is to develop a method that correctly associates a given special appeal with its most relevant repetitive themes. Our proposed method, GLARE, aims to associate a special appeal with relevant repetitive themes correctly. The method consists of two main steps: generating a summary of the appeal using Guided LexRank, followed by evaluating the similarity of that summary to predefined themes.This section details the methodology we adopted to generate and validate the proposed method to solve the problem in question. We start by presenting the issue at hand in a more formal way. Then, we proceed to detail the steps of our proposed method.

Consider a document corresponding to a special appeal and denoted by $r$. Also, consider a set of repetitive themes, which we denote by $\mathcal{T}$. The problem we consider corresponds to finding a subset $\mathcal{T}^\prime \subset \mathcal{T}$ of themes related to the content of document $r$. Furthermore, for each theme $t \in \mathcal{T}^\prime$, we want to produce a value $\sigma_{rt}$ that shows how related the themes are to the documents $r$ and $t$ and that allows you to order the elements in $\mathcal{T}^\prime \subset \mathcal{T}$. It is worth noting that an analyst human, when evaluating the elements in $\mathcal{T}^\prime$, will select at most one element that is the theme suitable for appeal $r$. 

\begin{figure}[htb]
\centering
\includegraphics[width=0.6\textwidth]{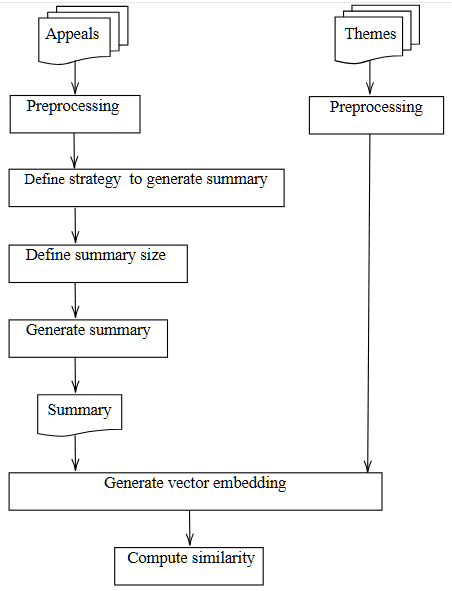}
\caption{Steps of the proposed method.}\label{fig:method:steps}
\end{figure}

The steps of the proposed method are listed below and will be detailed in this section. Figure~\ref{fig:method:steps} presents a schematic diagram of the defined steps.

\begin{enumerate}
     \item Process the textual content of each appeal $r$ and each theme $t$.
     
     \item Generate textual representation that synthesizes the content of appeal $r$ and vector embedding of this summary. Similarly generate vector embedding for the text of each theme $t \in \mathcal{T}$.
     
     \item Perform pairwise comparison between the textual representation of the appeal and each theme, producing the value $\sigma_{r^\prime t}$ associated with the level of similarity between the compared elements.
     
     \item Produce and sort the set $\mathcal{T}^\prime \subset \mathcal{T}$.
\end{enumerate}

\subsection{Preprocessing}
The special appeal is unstructured text, but it presents a set of paragraphs containing the document's core. In this method step, we select the section containing these paragraphs with the resource's essential information.

Once the document's core was identified, we applied a removal treatment to eliminate words, punctuation marks, and numerical patterns that are irrelevant to the task in question, which generate ``noise'' in identifying relevant information. We found and removed words considered stop words in Portuguese from the text. In general, articles, prepositions, and conjunctions. Due to the nature of the texts contained in the corpus, we use regular expressions to find and remove addresses and numbers from documents that name individuals and legal entities.

\subsection{Summarization Through Guided LexRank}
\label{subsubsec:summarization}

We adopted an extractive summarization approach to avoid possible hallucinations. In this way, when working with terms and structures that are, in fact, in the original text, we aim to preserve their semantics and avoid losses in the next stage of the method.

As described in Section~\ref{sec:background:lexrank}, LexRank represents text as a weighted graph, where nodes represent sentences and edges represent the similarity between sentences. The weights associated with nodes are related to the degree of centrality of each sentence. Considering the context of this work, when we send the text of a special appeal to LexRank, the algorithm outputs a summary, which is formed by the set of sentences that best represent the appeal. We should consider that a sentence not included in the summary might still be important for classifying the special appeal into a theme. LexRank can ignore a sentence because it calculates the centrality of each sentence without accounting for any external factors beyond the text itself.

~\citet{OTTERBACHER200942} present a version of the LexRank algorithm that considers bias based on a topic of interest. An approach is proposed to introduce bias that starts from the general LexRank equation and adds a priority term based on a weight function. According to the similarity that a given sentence in the text has with a previously defined topic, the sentence will receive a weight that will influence the composition of the final summary of the text.

We propose a change to the LexRank algorithm to guide the summarization task. With the proposed change, the decision for a sentence to compose the summary of the text is the result of a combination of two factors. One of the factors is internal to the text, that is, the degree of centrality of the sentence calculated in the original LexRank algorithm. The other factor is external to the text, calculated according to the similarity between each sentence to be summarized and the texts of the themes in this work. This way, in Guided LexRank, for each sentence s, the following steps are performed:
\begin{enumerate}
     \item Compute the degree centrality $\gamma_{s}$ in the same way as it is calculated in LexRank.
     \item Send each sentence $s$ as input to the algorithm BM25 targeting each theme $t \in T$. Store the highest score found when running BM25, $\sigma_{st}$.
     \item Calculate the combined score using two weighting factors (see Equation~\ref{eq:combined-score}).
\end{enumerate}

We propose a combined approach to compute the score, by using Equation~\ref{eq:combined-score}. In this equation, $\alpha$ and $\beta$ are arbitrary parameters. This way, the sentences selected to compose the text summary will be those with the highest combined score values. The number of chosen sentences, and so the size of the summary, stays limited by the ``size'' parameter of the original algorithm. Our proposed summarization approach, Guided LexRank, differs from Biased Lexrank~\citet{OTTERBACHER200942} in two aspects. First, Guided Lexrank uses a different algorithm to combine factors to weigh the relevance of a sentence. The second difference is that our method uses the BM25 algorithm to evaluate the importance of sentences in the text and sentences external to the text.

\begin{equation}
\label{eq:combined-score}
     \text{Combined score} = \alpha \times \gamma_{s}+\beta \times \sigma_{st}
\end{equation}

\subsection{Similarity computation}
\label{subsubsec:similarity:computation}
As described in Section~\ref{sec:methodology}, we possess the text/summary and its vector embedding at this stage. It is common practice to quantify the semantic similarity between texts by applying cosine similarity. However,~\citet{10.110.1145.3589335.3651526} warn against the blind use of this technique. Using empirical observations,~\citet{10.110.1145.3589335.3651526} demonstrate that the cosine similarity of learned embeddings can produce arbitrary results. Different similarity results can be obtained on the same set of data. These discrepant results are due to the various forms of training that the large language model (LLM) used to generate the vector embedding of the text can undergo. Considering the fact presented by~\citet{10.110.1145.3589335.3651526}, we sought to evaluate the similarity between special appeals and themes in two ways. One way is to use the text/summary as an input query for the BM25 algorithm, targeting the theme text. Another way is to calculate cosine similarity with the vector embedding of the appeal and themes.

\section{Experiments and results}
\label{sec:experiments}

The Section~\ref{sec:settings} details the experimental setup, including the computational environment and the Python libraries used to ensure reproducibility. The Section~\ref{sec:metrics} details the metrics used to evaluate the experiments.  Section~\ref{sec:experiments} evaluates the proposed model by varying preprocessing techniques, summarization methods, summary lengths, and similarity calculation approaches.  Section~\ref{sec:summarization:performance} explores the impact of the text summarization process, confirming that even with dimensionality reduction, summarization improves accuracy compared to using the full text. Section~\ref{sec:elasticsearch:baseline} compares our method with Elasticsearch baseline. Section~\ref{sec:supervised:baseline} compares our method with supervised models, showing that GLARE is competitive. This highlights the effectiveness of the approach.

\subsection{Experimental settings}
\label{sec:settings}
The experiments were run on an AMD EPYC 7452 32-Core 2.35GHz 52GB RAM server. The experiments were conducted using a set of Python libraries with specific versions. The libraries utilized include \textbf{argparse} version 1.1, \textbf{BERTopic} 0.16, \textbf{csv} 1.0, \textbf{LexRank} 0.1.0, \textbf{nltk} 3.6.7, \textbf{numpy} 1.22.4, \textbf{pandas} 2.1.4, \textbf{rank\_bm25} 0.2.2, \textbf{rank\_eval} 0.1.3, \textbf{re} 2.2.1, \textbf{sentence\_transformers} 2.2.2, and \textbf{torch} 1.12.1. These versions were selected to align with the methodology and ensure that the experimental results are reliable and reproducible.

\subsection{Evaluation metrics}
\label{sec:metrics}
As described in Section~\ref{sec:corpus}, each appeal that is part of the dataset has an associated theme defined by a human analyst. Having the list of themes and each appeal labelled by a human analyst allows us to evaluate the performance obtained with our method for assigning a theme to a special appeal. We use metrics from the context of information retrieval.

\subsubsection*{Recall}
\begin{equation}
        \label{eq:recall}
        \text{Recall at k}\ =\ \frac{\text{Relevant items in the k position ranking}}{\text{Total relevant items}}
\end{equation}

Considering that a human analyst classifies a special appeal into a theme, recall becomes the most relevant metric in this context, as, by definition, it evaluates the proportion with which a theme pertinent to the user is in the ranking of $k$ positions.

\subsubsection*{Mean Average Precision (MAP)}

Considering a list $\mathcal{L}$ having $k$ recommended elements in a query, the average precision is calculated considering the position in which a relevant item is found in the list. The MAP is the average of these average accuracies for all items.

\begin{equation}
    \label{precisionAtK}
    \text{Precision at k} = \frac{\text{Number of relevant items up to position k}}{k}
\end{equation}

\begin{equation}
    \label{averagePrecision}
    \text{Average Precision}= \frac{\sum_{k=1}^{k}(\text{Precision at k} * \text{Item relevance in k)}}{\text{Total number of relevant items}}
\end{equation}

\begin{equation}
    \label{map}
    \text{MAP at k} = \frac{\sum_{q=1}^{Q}\text{Average Precision for query q}}{Q}
\end{equation}

\subsubsection*{F1}

It is calculated by the harmonic mean between precision and recall.

\begin{equation}
    \label{f1}
    F1 = 2 \times \frac{\text{Precision} \times \text{Recall}}{\text{Precision} + \text{Recall}}
\end{equation}

\subsubsection*{Normalized Discounted Cumulative Gain (NDCG)}
\label{ndcg}

Like Precision at $k$, NDCG considers items' relevance and position in the ranking. According to Eq.~\ref{eq:dcg}, the accumulated gain is normalized to manage different relevance scales. In this equation, $\text{rel}_i$ is the degree of relevance of the element at position $i$.

\begin{equation}
    \label{eq:dcg}
    \text{DCG at k} = \sum_{i=1}^{k}\frac{2^{\text{rel}_i}-1}{\log_2(i+1)}
\end{equation}

\begin{equation}
    \text{IDCG at k is the ideal DCG, where items are ordered by relevance.}
\end{equation}

\begin{equation}
    \text{NDCG at k} = \frac{\text{DCG at k}}{\text{IDCG at k}}
\end{equation}

\begin{equation}
    \text{NDCG at k} = \frac{\sum_{q=1}^{Q}\text{NDCG for query q}}{Q}
\end{equation}

\subsection{Comparing pipeline alternatives}
\label{sec:comparative:experiments}
We conducted a series of comparative experiments to assess the performance of our legal case retrieval approach. These experiments aimed to evaluate the impact of various parameters and strategies on the method's effectiveness. By systematically varying preprocessing techniques, types of text representation, summary sizes, and similarity calculation methods, we sought to identify the optimal configuration for improving accuracy and relevance. This section details the experimental setup and the specific parameters tested to understand their influence on the model's performance. In summary, the parameters we vary in our comparative experiments consist of the following:
\begin{itemize}
     \item Pre-processing— We adopted removing or retaining terms from the special appeal as a decision parameter.
     \item Type of representation—In the step following preprocessing, we choose the type of representation that will be adopted later in the flow. We could keep the text or generate an extractive summary using the BERTopic, guided BERTopic, LexRank, or guided LexRank strategies.
     \item Summary size—We make variations in the sizes of the summaries generated using a parameter passed to the algorithm.
     \item Similarity calculation—At this stage, we possess the text/summary and its vector embedding. This way, we can evaluate the similarity of the appeal with the theme in two ways. One way is to use the text/summary as an input query for the BM25 algorithm targeting the theme text. Another way is to calculate cosine similarity with the vector embedding of the appeal and themes.
\end{itemize}

For each special appeal text sent as input into our method, the output was a list of 6 suggested themes considered most appropriate to the text. The best results are condensed in Tables~\ref{tab:removalBM25},~\ref{tab:removalCosine},~\ref{tab:noremovalBM25} and~\ref{tab:noremovalCosine} according to the variation parameters.

\begin{table}[htb]
\centering
\caption{Treatment: With the removal of terms. Similarity: BM25}
\begin{tabular}{lrrrr}
\hline
\textbf{Summarization technique} & \textbf{recall@6} & \textbf{F1-score} & \textbf{map@6} & \textbf{ndcg@6} \\ \hline
GLARE (Guided LexRank—15 sentences) & \textbf{0.7575} & \textbf{0.6268} & \textbf{0.5345} & \textbf{0.5902} \\
LexRank—60 sentences & 0.7215 & 0.6014 & 0.5157 & 0.5672 \\
BERTopic—55 words & 0.5696 & 0.4552 & 0.3791 & 0.4266 \\
Guided BERTopic—60 words & 0.5551 & 0.4525 & 0.3819 & 0.4253 \\
\hline
\end{tabular}
\label{tab:removalBM25}
\end{table}

\begin{table}[htb]
\centering
\caption{Treatment: With the removal of terms. Similarity: Cosine}
\begin{tabular}{lrrrr}
\hline
\textbf{Summarization technique} & \textbf{recall@6} & \textbf{F1-score} & \textbf{map@6} & \textbf{ndcg@6} \\ \hline
Guided LexRank—10 sentences        & 0.4704            & \textbf{0.3858}           & 0.3269        & \textbf{0.3626}         \\
LexRank—20 sentences        & 0.3543            & 0.2700           & 0.2181        & 0.2519         \\
BERTopic—45 words & \textbf{0.4792} & 0.3700 & \textbf{0.5157}        & 0.3454         \\
Guided BERTopic—45 words & 0.4612 & 0.3552           & 0.2887        & 0.3322         \\
\hline
\end{tabular}
\label{tab:removalCosine}
\end{table}

\begin{table}[htb]
\centering
\caption{Treatment: No removal of terms. Similarity: BM25}
\begin{tabular}{lrrrr}
\hline
\textbf{Summarization technique} & \textbf{recall@6} & \textbf{F1-score} & \textbf{map@6} & \textbf{ndcg@6} \\ \hline
LexRank—50 sentences        & 0.7302           & 0.5852           & 0.4883        & 0.5484         \\
GLARE (Guided LexRank—30 sentences)        & \textbf{0.7545}           & \textbf{0.6361}           & \textbf{0.5498}        & \textbf{0.6009}         \\
BERTopic—25 words & 0.5506           & 0.4452           & 0.3736         & 0.4180         \\ 
Guided BERTopic—60 words        & 0.5526            & 0.4500           & 0.3796        & 0.4228         \\
\hline
\end{tabular}
\label{tab:noremovalBM25}
\end{table}

\begin{table}[htb]
\centering
\caption{Treatment: No removal of terms. Similarity: Cosine}
\begin{tabular}{lrrrr}
\hline
\textbf{Summarization technique} & \textbf{recall@6} & \textbf{F1-score} & \textbf{map@6} & \textbf{ndcg@6} \\ \hline
LexRank—30 sentences        & 0.4176            & 0.3036           & 0.2385        & 0.2827         \\
Guided LexRank—10 sentences        & 0.3798            & 0.2804           & 0.2223        & 0.2610         \\
BERTopic—45 words        & \textbf{0.4638}           & \textbf{0.3605}           & \textbf{0.2949}        & \textbf{0.3373}\\
Guided BERTopic—45 words & 0.4616           & 0.3566           & 0.2905        & 0.3333         \\
\hline
\end{tabular}
\label{tab:noremovalCosine}
\end{table}

The best result obtained among all the experiments was a recall@6 equal to 0.75750. In summary, applying the method we propose, GLARE, which integrates Guided LexRank for summarization and BM25 for similarity evaluation, our method was able to suggest the correct theme for the appeal for approximately 76 percent of the records in our dataset.

The performance difference between the Guide LexRank method and BERTopic when using cosine similarity can be attributed to the nature of the summaries generated and the way similarity is calculated. While Guide LexRank, which is graph-based, prioritizes the structure and importance of sentences, it does not capture contextual semantic similarities as effectively, resulting in inferior performance compared to BERTopic when evaluated with cosine similarity. However, when BM25 is used for evaluation, which emphasizes exact term matching, Guide LexRank significantly outperforms BERTopic, highlighting its effectiveness in preserving key terms and relevant textual structure.

To better understand the results, we conducted further experiments to evaluate each summary generation strategy's behavior by varying the generated summary size. We also assessed the behavior of each summary generation strategy by removing terms considered irrelevant in the text. The goal was to find saturation points in the performance of the summarization strategies and the relationship with the variation parameters of the experiments.

\begin{figure}[htb]
    \centering
    \includegraphics[width=1.0\textwidth]{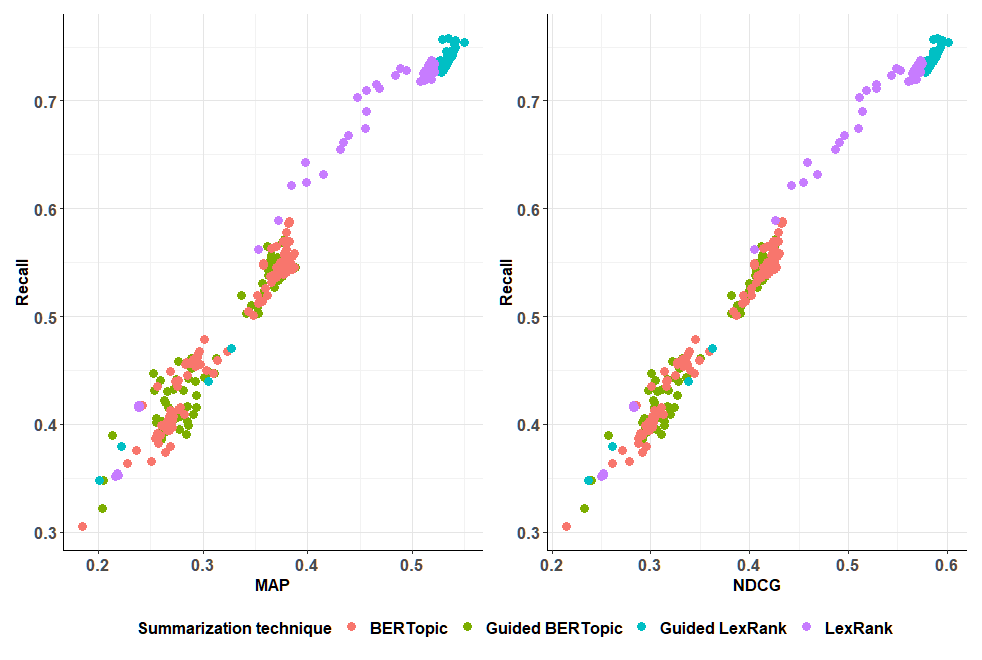}
    \caption{The quality of the ranking of suggestions, given by each summarization technique, can be measured by the MAP and NDCG metrics.}
    \label{fig:recall:mapNdcg}
\end{figure}

The best results are in the upper right quadrant in the scatter plot shown in Figure~\ref{fig:recall:mapNdcg}. This region stands for the results with the highest recall value associated with a ranking of better-quality suggestions, measured by the MAP and NDCG metrics. From what is shown in Figure~\ref{fig:recall:mapNdcg}, the best results were achieved with the algorithm we proposed, Guided LexRank.

\begin{figure}[htb]
    \centering
    \includegraphics[width=1.0\textwidth]{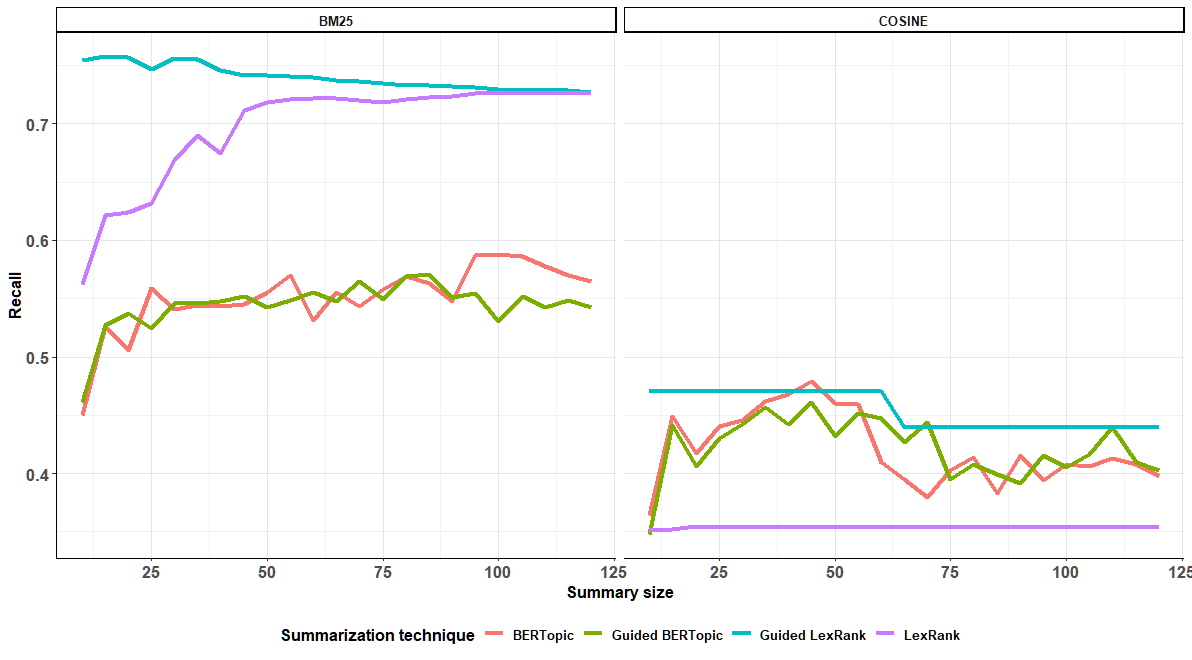}
    \caption{The model's performance is evaluated according to the number of sentences contained in the summary. Another aspect considered was the metric used to evaluate the similarity between the special appeal summary and the theme, whether by BM25 or cosine.}
    \label{fig:recall:summary}
\end{figure}

Based on the results shown in Figure~\ref{fig:recall:summary}, the method's performance depends on the proper combination of the summarization strategy and the similarity assessment method. The superior performance was clear when using the BM25 algorithm to evaluate the similarity between the special appeal summary and repetitive themes instead of using the cosine similarity measurement.

\begin{figure}[htb]
    \centering
    \includegraphics[width=1.0\textwidth]{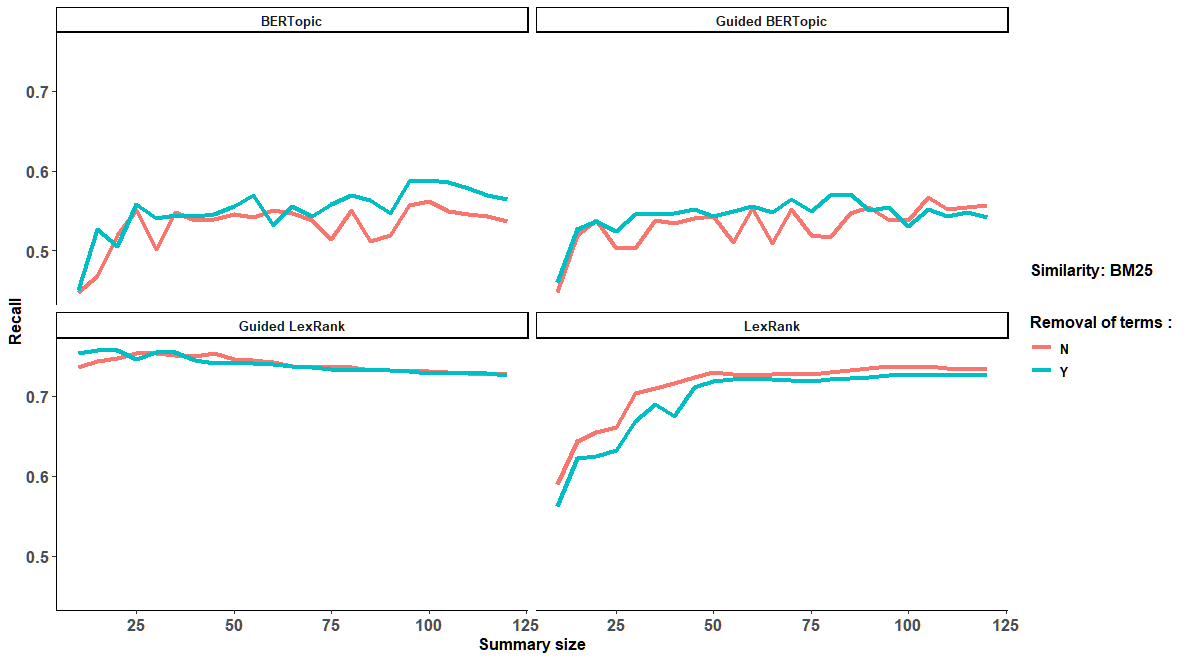}
    \caption{For each summarization technique applied, performance is evaluated by removing or not removing terms considered irrelevant in the original text of the special appeal. Other aspects considered were the similarity measured by BM25 and the number of sentences contained in the summary after applying the summarization techniques.}
    \label{fig:removal:bm25}
\end{figure}

Based on the results presented in Figure~\ref{fig:removal:bm25}, we cannot directly report the removal of terms from the text with improved method performance. However, the superior performance when combining similarity assessment using the BM25 algorithm with graph-based summarization techniques is notable.

\begin{figure}[htb]
    \centering
    \includegraphics[width=1.0\textwidth]{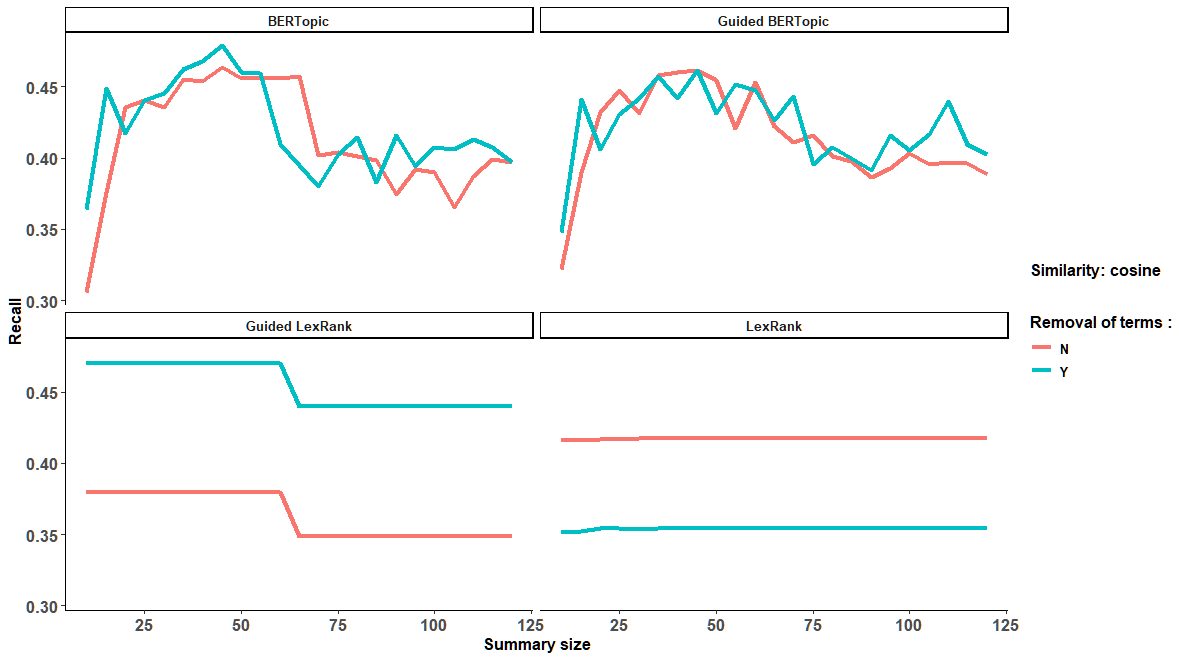}
    \caption{For each summarization technique applied, performance is evaluated by removing or not removing terms considered irrelevant in the original text of the special appeal. Other aspects considered were the similarity measured by cosine and the number of sentences contained in the summary after applying the summarization techniques.}
    \label{fig:removal:cosine}
\end{figure}

The results presented in Figure~\ref{fig:removal:cosine} highlight the effect of term removal, especially in graph-based summarization techniques. The guided LexRank technique benefits from removing irrelevant terms by applying cosine similarity. The opposite effect occurs when performing the same removal combined with the traditional LexRank technique. Also noteworthy is the constant performance in graph-based techniques, even with substantial variation in the number of sentences in the text summary.

When using the BM25 algorithm to generate a similarity measure, we seek an alternative to the traditional cosine similarity measure. As described in Section~\ref{subsubsec:similarity:computation}, the parameters and training data used by a model that generates the vector embedding of a text directly affect the cosine similarity calculation. Based on the results obtained in the experiments, we confirmed the viability and superiority of the BM25, about cosine, in the task performed. We verified the effectiveness of the proposed method for finding the theme of a special appeal by providing the appeal text and a list of repetitive themes. Combining the graph-based algorithm we propose, guided LexRank, with the BM25 algorithm has produced effective gains over other approaches we have analyzed.

\subsection{Impact of text summarization on performance}
\label{sec:summarization:performance}

\citet{DBLP:books.sp.Aggarwal22} claims that some dimensionality reduction process can change a document's representation. Although it might seem that dimensionality reduction loses information, it is often possible to choose a representation dimensionality in which the semantic knowledge in the corpus is kept and most of the noise is reduced.

The objective of the summarization step proposed in our method (Section~\ref{subsubsec:summarization}) is to reduce noise to improve the quality of the representation and consequently facilitate the process of classifying the special appeal in a theme. In all experiments, we obtained a recall greater than zero, showing that using the text summary effectively to classify a special appeal in a theme is effective. However, we also wanted to assess whether there was a gain when using the summary instead of the full text. To evaluate this aspect, we included two cases in our experiments in which we used the full text, as described below.

\begin{itemize}
    \item Assessment of similarity between the special appeal's full text and each theme's text using BM25.
    \item Similarity assessment between the vector embedding of the full text of the special appeal and the vector embedding of each theme using cosine.
\end{itemize}

Table~\ref{tab:fulltext} presents the comparative analysis of full text versus summarized text. In this table, we use the following mnemonics to represent each alternative: FT (i.e., full text is used), GLARE15 (i.e., Guided LexRank- summary with 15 sentences), FTE (i.e., full text embedding), and GLR10 (i.e., Guided LexRank-summary with 10 sentences-embedding). According to these results, we conclude that it is possible to suppress terms from a text and still obtain a result that is superior to the result obtained using the full text.

\begin{table}[htb]
\centering
\caption{Comparative analysis of full text vs. summarized text}
\begin{tabular}{lrrrrr}
\hline
\textbf{Representation type} & \textbf{similarity} & \textbf{recall@6} & \textbf{F1-score} & \textbf{map@6} & \textbf{ndcg@6} \\ \hline
FT                    &  BM25            & 0.7146           & 0.6041           & 0.5233        & 0.5714         \\
GLARE15 & BM25  & \textbf{0.7575} & \textbf{0.6268} & \textbf{0.5345} & \textbf{0.5902} \\
FTE        &  Cosine          & 0.1363           & 0.1073           & 0.0884         & 0.0999         \\
GLR10 &  Cosine          & 0.4704            & 0.3858           & 0.3269        & 0.3626        \\
\hline
\end{tabular}
\label{tab:fulltext}
\end{table}

\subsection{Comparison with Elasticsearch Baseline}
\label{sec:elasticsearch:baseline}

As previously mentioned in Section~\ref{sec:introduction}, Elasticsearch was identified as one of the baseline models for comparison in this study. The performance of the proposed method, GLARE, was evaluated against this baseline to assess its effectiveness in classifying special resources. 

The results, detailed in Table~\ref{tab:elasticsearch}, highlight the significant improvements achieved with GLARE. Our method demonstrated superior performance across all evaluated metrics. These results underscore the effectiveness of the proposed approach in identifying relevant themes for special resources, surpassing the performance of the Elasticsearch baseline.

\begin{table}[htb]
\centering
\caption{Performance Comparison between Baseline Elasticsearch and GLARE.}
\begin{tabular}{lrrrr}
\hline
\textbf{Representation type} & \textbf{recall@6} & \textbf{F1-score} & \textbf{map@6} & \textbf{ndcg@6} \\ \hline
GLARE & \textbf{0.7575} & \textbf{0.6268} & \textbf{0.5345} & \textbf{0.5902} \\
Baseline-Elasticsearch&0.3491&0.3084&0.3275&0.3182 \\
\hline
\end{tabular}
\label{tab:elasticsearch}
\end{table}

\subsection{Comparison with Supervised Learning Baselines}
\label{sec:supervised:baseline}

We framed the problem of assigning a ranked list of themes to a given appeal as a supervised classification problem by treating themes as labels. Each appeal in the corpus is associated with a specific theme, and these themes serve as target labels for the classification models. By framing the problem in this way, we can train supervised models to learn the relationship between the content of documents and their corresponding themes.

To obtain a ranking of the themes, similar to the unsupervised approach, we exploit the probabilities generated by the supervised models. We use XGBoost~\citep{Chen_2016} and Logistic Regression~\citep{scikit-learn} models, both configured in a one-vs-rest (OvR) scheme, where each model independently learns to distinguish one theme from all others. After training, the models return a probability associated with each theme for each document. These probabilities are then sorted in descending order to produce a ranking, providing an output consistent with our unsupervised method's proposal.

\subsubsection*{Cross-Validation}
\label{sec:crossvalidation}

The cross-validation technique divides the labeled data into $k$ equal segments. One of the $k$ segments is used for testing, and the remaining ($k-1$) segments are used for training. This process is repeated $k$ times by using each of the $k$ segments as the test set~\citep{DBLP:books.sp.Aggarwal22}.
In both supervised models, we used a ten-fold stratified cross-validation technique, ensuring that each fold had the same proportion of classes as the original dataset.

\subsubsection*{Minority Classes Resampling}

As shown in Figure ~\ref{fig:histogram:themes}, the corpus used in our project has some class imbalance. To address the class imbalance, we conducted two experiments with the supervised models: one without applying any balancing technique and the other using a resampling method available in the scikit-learn library. The resampling technique artificially increases the number of minority class examples in the training set, equalizing the number of examples across classes. Applying this technique helped us evaluate the impact of balancing on the model performance.

\subsubsection*{Model Training and Calibration}

We used the XGBoost and Logistic Regression classifiers in a one-vs-rest (OvR) scheme, allowing multi-class classification. Models' probability outputs were calibrated using a resource available on Sci-Kit Learn with the sigmoid method, based on~\citet{10.1145/775047.775151}.

With the techniques used, we aimed to build robust models that could serve as a reliable baseline for comparison with our proposed unsupervised approach. The metrics in Table~\ref{tab:supervised} compare our model with the supervised models. The data related to the supervised models are presented in the format \textit{arithmetic mean $\pm$ standard deviation}, since they refer to the result obtained by applying cross-validation in 10 folds.

\begin{table}[htb]
\centering
\resizebox{\textwidth}{!}{%
\begin{tabular}{lcrrrr}
\hline
\textbf{Model} &\textbf{Resampling} & \textbf{recall@6} & \textbf{F1-score} & \textbf{map@6} & \textbf{ndcg@6} \\ \hline
LogReg & Yes & 0.9494$\pm$0.0050 & 0.6835$\pm$0.0089& 0.1329$\pm$0.0007& 0.9114$\pm$0.0054  \\ 
LogReg & No & \textbf{0.9543$\pm$0.0056} & 0.6712$\pm$0.0082& 0.1365$\pm$0.0008& \textbf{0.9242$\pm$0.0043}  \\ 
XGBoost & Yes & 0.8712$\pm$0.0103 & 0.7215$\pm$0.0100& 0.1313$\pm$0.0013& 0.8553$\pm$0.0080  \\ 
XGBoost & No & 0.9065$\pm$0.0086 & \textbf{0.7447$\pm$0.0090}& 0.1364$\pm$0.0014& 0.8861$\pm$0.8861  \\ 
GLARE & n/a   & 0.7575 & 0.6268 & \textbf{0.5345} & 0.5902 \\ \hline

\end{tabular}%
}
\caption{Comparative performance: unsupervised vs. supervised algorithms (XGBoost: Extreme Gradient Boosting. LogReg: Logistic Regression).}
\label{tab:supervised}
\end{table}

\subsubsection*{Performance in classes with few examples}

As reported in Section~\ref{sec:introduction}, it is common for a court's database to have a few special appeals labeled on some specific themes. In certain scenarios, there is a possibility that the appeal recently submitted to the court presents an unprecedented issue to the Judiciary. These are important situations that the human expert deals with when analyzing a special appeal. Therefore, they are the premises on which we developed our method to classify the special appeal of a theme.

To evaluate the performance of the supervised models in classes with few examples, we conducted an experiment in which we extracted two disjoint subsets of the original corpus obeying the following rules:
\begin{itemize}
    \item The model training set has 80\% of the original corpus.
    \item The model test set has 20\% of the original corpus.
    \item The test set mostly comprises classes with the lowest representation in the original corpus.
    \item The training set is guaranteed to have at least one example of each class from the original corpus.
    \item Respecting the imposed rules, data insertion into the training and test sets occurs randomly.
\end{itemize}

\begin{figure}[htb]
    \centering
    \includegraphics[width=1.0\textwidth]{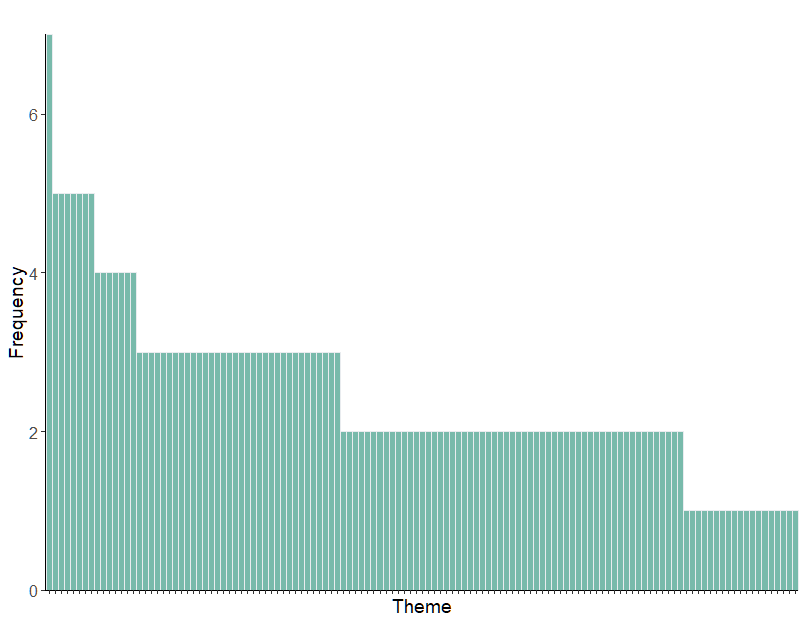}
    \caption[histogram]{Distribution of themes in the data set for training the supervised model.}
    \label{fig:histogram:minority}
\end{figure}

The histogram in Figure~\ref{fig:histogram:minority} shows the frequency of these minority classes in the supervised model training set.

The best hyperparameter configuration for the model was identified by applying cross-validation to the training set. After that, the model was trained on the entire training set (Section~\ref{sec:crossvalidation}). This final model was applied to the previously separated test set (20\% of the original corpus data). Since the test set was not used at any point during training or validation, this provides us with an unbiased assessment of the model's performance.

During this experiment, we saw that the supervised models performed poorly. This result was particularly surprising when compared to the performance of our unsupervised method running on the same data set. Our model made correct theme attribution in 72\% of cases, as seen in Table~\ref{tab:fewExamples}.

\begin{table}[htb]
\centering
\small
\caption{Comparative Performance: unsupervised vs. supervised algorithm with minority classes}
\begin{tabular}{lcrrrr}
\hline
\textbf{Model} &\textbf{Resampling} & \textbf{recall@6} & \textbf{F1-score} & \textbf{map@6} & \textbf{ndcg@6} \\ \hline
Logistic regression & Yes & 0.5442 & 0.3903& 0.0696& 0.4818  \\ 
Logistic regression & No  & 0.6064 & 0.3223 & 0.0646& 0.4920  \\ 
XGBoost & Yes & 0.2241 & 0.2323& 0.0322& 0.2075  \\ 
XGBoost & No & 0.2931 & 0.2412& 0.0375& 0.2548  \\ 
GLARE & Not applicable  & \textbf{0.7200} & \textbf{0.6059} & \textbf{0.5230} & \textbf{0.5717} \\

\hline
\end{tabular}
\label{tab:fewExamples}
\end{table}

\subsubsection{Assessing Model Generalization on Unrepresented Legal Themes in the Training Data}

In another experiment, we examined an extreme scenario where special appeal documents in the test set align with topics without documents in the training set. Through this experiment, we replicated the situation where a human expert encounters a special appeal that does not resemble any content in a historical database. Even without another previously classified reference document, the expert should be able to analyze the document and classify it into a topic from the list of repetitive topics (Table~\ref{tab:zeroExamples}).

\begin{table}[htb]
\centering
\small
\caption{Comparative Performance: Unsupervised Algorithm vs. Supervised logistic regression, with classes without representatives in the training set.}
\begin{tabular}{lcrrrr}
\hline
\textbf{Model} &\textbf{Resampling} & \textbf{recall@6} & \textbf{F1-score} & \textbf{map@6} & \textbf{ndcg@6} \\ \hline
Logistic regression & Yes & 0.0000 & 0.0000& 0.0000& 0.0000  \\ 
Logistic regression & No & 0.0000 & 0.0000& 0.0000& 0.0000  \\ 
XGBoost & Yes & 0.0000 & 0.0000& 0.0000& 0.0000  \\ 
XGBoost & No & 0.0000 & 0.0000& 0.0000& 0.0000  \\ 
GLARE & Not applicable  & \textbf{0.6903} & \textbf{0.5917}& \textbf{0.5177}& \textbf{0.5598}  \\

\hline
\end{tabular}
\label{tab:zeroExamples}
\end{table}

The results showed that the supervised models struggled with generalization. They were unable to make correct assessments on classes for which they were not trained.
It was evident that a substantial volume of labeled data is essential for the supervised model to learn representative patterns. In contrast, our unsupervised model proved to be effective in the scenario analyzed. In approximately 70\% of the cases, the model correctly identified the theme of the special appeal.

\section{Conclusion}
\label{sec:conclusion}

In this work we investigated the problem of associating a special appeal to a ranked list of themes. We developed our method based on the hypothesis that, given the textual content of a special appeal, it would be possible to generate a summary that preserved the essential information of that text and that, from this summary, we could evaluate the corresponding special appeal similarities with themes.

Through several experimental results, we confirmed the effectiveness of our method, especially in a scenario where there are few document examples for a given class. We also showed the feasibility of using the BM25 algorithm to analyze the similarity between texts, evaluating it as a more effective alternative to the traditional cosine similarity assessment. Our method, GLARE, combines the Guided LexRank for summarization and BM25 for similarity evaluation, offering an unsupervised and effective solution.

The results obtained confirmed the quality of the summaries generated from graph-based algorithms. The experiments also confirmed that the modification we implemented in the LexRank algorithm, generating a guided version of the algorithm, positively impacted the results.

Our method does not consider any specific linguistic structure and can be applied in contexts beyond theme attribution of special appeals and even the legal domain. As it is an unsupervised machine learning approach, it does not require data for training like supervised learning. It proved an effective and versatile solution suitable for other problems not explored in this project.

Since our method's performance was much higher than the baseline in scenarios with minority classes, a promising direction for future work would be to develop a hybrid system that combines supervised and unsupervised learning, creating a more robust and efficient system.

Additionally, future work could explore the integration of Large Language Models (LLMs) as part of the preprocessing pipeline in our method. LLMs, such as GPT-4, have shown significant potential in enhancing data preprocessing tasks, including text normalization, error detection, and data imputation~\citep{zhang2023largelanguagemodelsdata}. By leveraging LLMs, it is possible to improve the quality and consistency of the input data, leading to even better performance and broader applicability of our method across various legal and non-legal domains.

\subsection*{Computer Code Availability}
All of the experiments described in this paper were developed with Python 3.10 programming language. For reproducibility purposes, our source code is publicly available at \url{https://github.com/AILAB-CEFET-RJ/r2t}.

\subsection*{Data Availability}

In this paper, datasets in CSV format were used and can be downloaded at Zenodo, an open-source online data repository: \url{http://doi.org/10.5281/zenodo.13696090}.

\subsection*{Acknowledgment}
The authors thank CNPq, CAPES, FAPERJ, and CEFET/RJ for partially funding this research.

\bibliography{bibliography}

\end{document}